\documentclass[aps,10pt,prl,twocolumn,a4paper,footinbib,superscriptaddress,floatfix]{revtex4}

\usepackage{amsmath,amssymb}
\usepackage{hyperref}
\usepackage{bbm}
\usepackage{slashed}
\usepackage{graphicx}
\usepackage{mathptmx}
\usepackage{xcolor}
\usepackage{natbib}

\usepackage{verbatim}

\newcommand{\bw}{\begin{widetext}}
\newcommand{\ew}{\end{widetext}}

\newcommand{\be}{\begin{equation}}
\newcommand{\ee}{\end{equation}}
\newcommand{\bestar}{\begin{equation*}}
\newcommand{\eestar}{\end{equation*}}

\newcommand{\bi}{\begin{itemize}}
\newcommand{\ei}{\end{itemize}}
\newcommand{\bea}{\begin{eqnarray}}
\newcommand{\eea}{\end{eqnarray}}

\newcommand{\hbo}{\hbox to 1 true cm {\hfill } }

\newcommand{\Eins}{\mathbbmss{1}}

\newcommand{\ud}{\mathrm{d}}

\newcommand{\LCm}{{\scriptscriptstyle -}} 
\newcommand{\LCp}{{\scriptscriptstyle +}}
\newcommand{\LCpm}{{\scriptscriptstyle \pm}}

\newcommand{\LCperp}{{\scriptscriptstyle \perp}}

\renewcommand{\sp}[2]{#1\!\cdot\!#2}

\begin{document}
\title{Trident pair production in strong laser pulses}
\author{Anton Ilderton}\email[]{anton.ilderton@physics.umu.se}
\affiliation{Department of Physics, Ume\aa\ University, SE-901 87 Ume\aa, Sweden}

\begin{abstract}
\noindent  We calculate the trident pair production amplitude in a strong laser background. We allow for finite pulse duration, while still treating the laser fields nonperturbatively in strong-field QED. Our approach reveals explicitly the individual contributions of the one-step and two-step processes. We also expose the role gauge invariance plays in the amplitudes and discuss the relation between our results and the optical theorem.
\end{abstract}

\maketitle
Electron-positron pair production, at the focus of an intense laser, is currently a topic of considerable interest due to the development of extreme light sources such as ELI \cite{ELI}. As is typical for particle scattering experiments, many different processes may contribute to the final yield of pairs. It is important to be able to judge the relative importance of these processes, and distinguish their contributions from each other. In the SLAC E144 experiment \cite{Bamber:1999zt}, pairs were produced by colliding a (low intensity) laser with the SLAC electron beam; high energy photons radiated by the electrons then combined with photons in the laser to produce pairs.  In that experiment, a process called `trident' may also have produced pairs. However, its contribution could only be evaluated approximately, since no exact expression for the trident amplitude was available, nor has one been given to date \cite{Bula:1997eh}. Accounting theoretically for all relevant effects in modern laser experiments is a formidable challenge. As well as contributions from processes such as trident (along with, for example, vacuum pair production \cite{Dunne:2008kc} and cascades \cite{Elkina:2010up}), one would like to include effects due to the properties of the laser itself, such as their ultra-high intensity (currently $10^{22}$ W/cm$^2$), ultra-short  duration (measured in femtoseconds or even attoseconds) and tight focus (focal diameter of the order of $10$ microns).

In this Letter, we present the first complete calculation of the trident process. We include high-intensity effects by treating the laser field nonperturbatively and exactly. We include finite size effects by allowing finite pulse duration. We begin by describing the essential features of the trident process, modelling the laser as a null field and using Volkov solutions to describe the scattered particles. Key aspects of existing calculations are discussed, and shown to be unphysical due to violations of the Ward identities of QED. We then calculate the scattering amplitude, emphasising its physical content, and relate our results to the optical theorem. We use gauge invariance to simplify the final expression for the emission rate, and discuss approximations useful to the high intensity regime.  

\begin{figure}[t!]
\includegraphics[width=0.55\columnwidth]{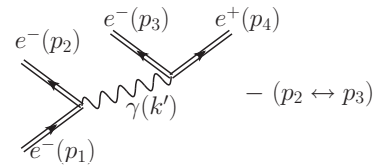}
\caption{\label{trident:fig} The trident process. Double lines represent the fermion propagator in the background field. The second term accounts for the exchange of indistinguishable fermions.}
\end{figure}

We consider an electron, incident upon a laser field, emitting a photon. This photon then combines with photons in the laser to produce an electron-positron pair. The relevant Feynman diagram is shown in Fig.~\ref{trident:fig}. Double lines represent the fermion propagator in the background laser field. We take this background to be a plane wave, with field strength $F_{\mu\nu}\equiv F_{\mu\nu}(\sp{k}{x})$, where the momentum $k_\mu$ is lightlike, so $k^2=0$. The laser frequency is $\omega = |\mathbf k|$. Our treatment holds for arbitrary $\sp{k}{x}$ dependence: in particular, we have finite pulse duration when $F_{\mu\nu}$ vanishes, or goes rapidly to zero, outside of some $\sp{k}{x}$ range.

Fig.~\ref{trident:fig} actually describes two processes. The first is `one step', in which the intermediate photon is virtual: this is the process traditionally referred to as trident. The second process is `two step', in which a real photon is scattered from the incoming electron (allowed in a background field via nonlinear Compton scattering \cite{Harvey:2009ry}), and this real photon then goes on to create a pair via stimulated pair production~\cite{Heinzl:2010vg}. This was the process of interest in the SLAC experiment. Since the  Feynman diagram in Fig.~\ref{trident:fig} makes no distinction between these processes, we will refer to the full diagram as `trident', and use the notions of one- and two-step processes to distinguish the two contributions.
The trident S-Matrix element is easily written down, using strong-field QED, in terms of the photon propagator $G_{\mu\nu}$ and Volkov solutions $\psi_j$ carrying the asymptotic momenta $p_j$ in Fig.~\ref{trident:fig} \cite{Volkov:1935} : 
\begin{align}
	\nonumber S_{fi} = e^2\!\int\!\ud^4x\int\! \ud^4 y\ \bar\psi_{2}(x) \gamma^\mu &\psi_{1}(x) G_{\mu\nu}(x-y)\bar\psi_{3}(y) \gamma^\nu \psi_{4}^\LCp(y) \\
	\label{begin} &- (p_2\leftrightarrow p_3) \;.
\end{align}
The Volkov solutions also carry a dependence on $\sp{k}{x}$, and $\sp{k}{y}$, which describes the ``dressing" of the particles by the background. One trades this dependence, via Fourier transform, for new variables $r$ and $s$: in the periodic plane wave case, the dressing is responsible for the shifted electron mass, and the Fourier transform reduces the S-Matrix element to a discrete sum over processes involving different numbers of `laser photons' \cite{Nikishov:1964zza, Nikishov:1964zz}. Our continuous $r$ and $s$ are the analogues of photon number for pulsed plane waves, see \cite{Heinzl:2010vg}. After this, the $x$ and $y$ integrals trivially yield two delta functions conserving four-momentum. One of these fixes the intermediate photon momentum $k'$, and we have, writing $\delta p \equiv p_1-p_2$ and $p_\text{out}\equiv p_2+p_3+p_4$ from here on,
\bea \label{S1}
	\nonumber S_{fi} = & (2\pi)^2e^2\displaystyle\int\! \ud r\!\int\!\ud s\  M(r,s) \delta^4(p_\text{out} -p_1 - r k - s k) \\
	& \times \displaystyle\frac{1}{k'^2+i\epsilon}\bigg|_{k' = \delta p + r k}  - (2\leftrightarrow 3)\;.
\eea
The expression for $M$ will be given soon: it is lengthy and we do not need it yet. The Fourier transformed (Feynman gauge) photon propagator is $g_{\mu\nu} / (k'^2+i\epsilon)$. The remaining delta function, above, sees the asymptotic momenta and hence the free electron mass, not the shifted mass of the infinite plane wave calculations; this is a consequence of allowing finite pulse duration, just as is observed in \cite{Heinzl:2010vg} for stimulated pair production and in \cite{Mackenroth:2010jr} for nonlinear Compton scattering in pulsed fields. From the delta function in (\ref{S1}) it becomes clear that $r$ and $s$ parameterise the energy-momentum taken from the laser.

In the current literature, a divergence has been identified in (\ref{begin}), arising from the pole in the photon propagator in the two-step process, i.e.\ when enough energy is  taken from the background to put the photon on-shell and $k'^2=0$. This divergence was attributed to the infinite temporo-spatial extent of the (periodic plane wave) background considered. Based on this, it was suggested that the divergence could be dealt with by modifying the photon propagator according to \cite{Hu:2010ye}
\be\label{wrong}
	\frac{1}{k'^2}\overset{!}{\to} \frac{1}{k'^2+2i|k'_0|/T}\;,
\ee
which damps the propagator outside of a chosen, fixed, temporal range $T$, in order to model finite pulse duration. However, it is well known that the structure of propagators and vertices in QED is severely restricted by gauge invariance, violations of which are measured by the Ward identities \cite{Abbott:1981ke}. The prescription (\ref{wrong}) manifestly violates the Ward identity \cite{gauge}
\be
	k'^\mu\langle\, A_\mu(k') A_\nu(q)\,\rangle = \frac{k'_\nu}{k'^2}\delta^4(k'+q)\;,
\ee
which is sufficient to reject (\ref{wrong}). We briefly state some additional physical reasons. First, the origin of the divergence cannot be the infinite extent of the background. Any background which permits the two-step process, producing an on-shell intermediate photon, will admit the $k'^2=0$ divergence, independent of the background's spacetime support (even if the theory contains a natural analogue of the cutoff scale $T$, for example). Secondly, there is also no reason why any chosen $T$ should be preferred by the theory over any other. The prescription (\ref{wrong}) also breaks manifest Lorentz invariance. While this may not be considered too serious, given that the background field already introduces preferred directions in spacetime, these directions are lightlike, not timelike as (\ref{wrong}) implies. Using (\ref{wrong}) also gives a $T$ dependent correction to the divergence-free one-step process, and leaves unphysical factors of $T$ in the weak field limit where all background effects should become negligible. It is therefore difficult to ascribe a physical meaning to results following from (\ref{wrong}).

The resolution of the above problem is simple: in a complete treatment, there is no divergence. It appears only when one neglects the pole prescription  already contained in Feynman propagators. The photon propagator is really $1/(k'^2+i\epsilon)$ and one must remember to take the infinitesimal $\epsilon\to 0^\LCp$.  To show there is no divergence, we return to (\ref{S1}), where the propagator is evaluated at $k' = \delta p+r k$. This can certainly go on-shell, so $k'^2=0$, for certain $r$. This behaviour may be exposed by recalling the distributional result:
\be\label{right}
	\frac{1}{k'^2+i\epsilon} \overset{\epsilon\to 0^\LCp}{=} -i\pi\delta(k'^2) + \mathcal{P}\frac{1}{k'^2} \;,
\ee
which clearly separates contributions from, respectively, real (on-shell) and virtual (off-shell) photons. We will reaffirm this statement below. Here, we perform the essential step, inserting (\ref{right}) into (\ref{S1}). In the first term, the delta-function is $\delta(k'^2) =  \delta(2r \sp{k}{\delta p} + \delta p^2)$ and is therefore eliminated by performing the $r$ integral. This fixes
\be
	r  = -\frac{\delta p^2}{2\sp{k}{\delta p}} \equiv r' \;,
\ee
so the propagator's pole fixes, quite naturally, the energy transfer from the laser such that the photon goes on-shell, i.e.\ such that the two-step process occurs. There is no divergence here, nor any need to deform the theory along the lines of (\ref{wrong}). (Note, the denominator of $r'$ is nonzero because $\sp{k}{\delta p} = \sp{k}{(p_3+p_4)}>0$, using the momentum conservation law in (\ref{S1}).)  To proceed, one eliminates the $s$ integral using one component of the delta function in (\ref{S1}). This is easily done in lightfront co-ordinates $x^\LCpm := x^0\pm x^3$, $x^\LCperp := \{x^1,x^2\}$, and $q_\LCpm=(q_0\pm q_3)/2$ for momenta. Since the laser momentum $k_\mu$ is lightlike, we may choose $k_\LCp$ to be its only nonzero component. Hence, $s$ appears only in the $q_\LCp$ component of the delta function, which eliminates the $s$ integral and sets
\be\label{sr}
	s  = \frac{p_\text{out}^2-m^2}{2\sp{k}{p_1}} - r \equiv s_r.
\ee
Three components of the delta function remain, conserving momentum in the $-$ and $\perp$ directions, transverse to the laser. Writing $\delta^{\mathsf{lf}}(q) \equiv \delta^2(q_\LCperp)\delta(q_\LCm)/k_\LCp$, the S-Matrix element is
\begin{align}
\label{full}
	 &S_{fi}= 2 e^2\pi^2\,\delta^\mathsf{lf}(p_\text{out} - p_1) \times \\
	\nonumber &\bigg[\displaystyle \frac{-i\pi}{2\sp{k}{\delta p}}M(r',s_{r'}) + \int\!\ud r\ M(r,s_r)\ \mathcal{P}\frac{1}{(\delta p + r k)^2} -(2\leftrightarrow 3)\bigg]\;.
\end{align}
This completes the calculation of $S_{fi}$. It should be clear from (\ref{right}) that the off/on-shell parts of (\ref{full}) correspond precisely to the one/two-step processes. Nevertheless, let us confirm this, which will also serve as a check on $S_{fi}$ before moving on to the cross section. Consider first the one-step process, with overall momentum conservation given by the explicit delta-function in (\ref{S1}). Squaring the argument of the delta function, one finds that $r+s >{4m^2}/{(\sp{k}{p_1})}$, which states that the total incoming energy (that of the initial electron and that taken from the laser), must be sufficient to produce three particles of rest mass $m$.  This is the only constraint in $S_{fi}$ if the photon is off-shell, $k'^2\not=0$, and we pick up the principal value term. This is the `trident' contribution in the old nomenclature. (Again, there is no dependence on the shifted mass found in a periodic plane wave.) Consider now the two-step process, which has two momentum conservation relations
\be\begin{cases}
	p_1 + r k = p_2 + k' & \text{nonlinear Compton scattering}\;,\\
	k' + s k = p_3 + p_4 & \text{stimulated pair production}\;.
\end{cases}\ee
Squaring, one finds that, for the individual processes to occur, 
\be\label{kinematics}
	r>0\quad\text{and}\quad s> \frac{2m^2}{\sp{k}{k'}}.
\ee
Using the overall delta function in (\ref{S1}), it is straightforward to show that the fixed parameters $r'$ and $s_{r'}$ which appear in our on-shell part obey  $r'>0$ and $r_{s'}> 2m^2/\sp{k}{\delta p}$. Recalling that $\sp{k}{\delta p} = \sp{k}{k'}$ evaluated on shell, we recover (\ref{kinematics}). Our solution thus contains the correct kinematics of both the one- and two-step processes, which are described precisely by the off- and on-shell parts of the decomposition (\ref{right}). Thus, our solution allows the contributions of these processes to be individually calculated and compared.

The identity (\ref{right}) also corresponds to a split into real and imaginary parts. From the optical theorem for scattering amplitudes (that is, unitarity of the S-Matrix), one expects the appearance of imaginary, or absorptive, parts to correspond to the excitation of real rather than virtual intermediate states \cite{Cutting, Branch}. Indeed, we have seen that the imaginary part of (\ref{right}) corresponds to the intermediate photon becoming real. The existence of this imaginary part is entirely due to the dressing of the fermions by the background: in cutting language, the dressing allows one to perform a cut through Fig.~\ref{trident:fig} which yields physical, non-zero scattering amplitudes; those for nonlinear Compton scattering and stimulated pair production. (The trident amplitude itself may be obtained by cutting the two-loop fermion propagator \cite{Ritus:1972nf}.) Our result therefore appears to contain a rather novel example of the optical theorem at  tree level, made possible by the background.  We hope to investigate this further in the future.

In the remainder of this Letter we give the complete calculation of the emission rate. We will also show how gauge invariance simplifies the final results. 

Our background field has potential $A_{\mu}(\sp{k}{x}) = f_j(\sp{k}{x}) a^j$, where $j$ is summed over the transverse directions. The polarisation vectors obey $\sp{a^i}{k}=0$, $\sp{a^i}{a^j}=-m^2a^2/e^2\delta^{ij}$ which defines an invariant, dimensionless amplitude $a$. From the corresponding Volkov solutions, we define
\be\begin{split}\label{PhaseSpin}
	J(p,b,c) &= -\frac{1}{2\sp{k}{p}}\int\limits_b^c\! \ud \phi\ 2e A(\phi)\!\cdot\!p - e^2 A^2(\phi) \;,  \\
	S(p,\sp{k}{x}) &= \Eins_4 + \frac{e}{2\sp{k}{p}} \slashed{A}(\sp{k}{x})\slashed{k}\;.
\end{split}
\ee
These are combined into `nonlinear Compton', $\Gamma^\mu$, and `pair production', $\Delta^\mu$, parts as follows (we denote the reverse-ordered $S$ by $\hat{S}$):
\be\label{zung}\begin{split}
	\Gamma^\mu(\sp{k}{x}) := &{\bar u}_{p_2}\hat{S}(p_2,\sp{k}{x}) \gamma^\mu S(p_1,\sp{k}{x})u_{p_1} \\
	&\times \exp[i J(p_2,\sp{k}{x},\infty)+ i J(p_1,-\infty,\sp{k}{x})] \;, \\
	\Delta^\mu(\sp{k}{y}) := &{\bar u}_{p_3}\hat{S}(p_3,\sp{k}{y}) \gamma^\mu S(-p_4,\sp{k}{y})v_{p_4} \\
	&\times \exp[iJ(p_3,\sp{k}{y},\infty)+ iJ(-p_4,\infty,\sp{k}{y})]\;.
\end{split}
\ee
The limits in $J$ are prescribed by the LSZ reduction formula for the trident amplitude.  The product of the Fourier transforms of these parts gives us the amplitude $M(r,s)$ from above,  
\begin{eqnarray}
	\nonumber M(r,s) &=& \displaystyle\int\!\ud \phi\ \Gamma^\mu(\phi)e^{ir  \phi} \int\ud \varphi\ \Delta_\mu(\varphi) e^{i s \varphi} -(2\leftrightarrow 3) \\
	&\equiv& {\tilde\Gamma}^\mu(r){\tilde\Delta}_\mu(s) -(2\leftrightarrow 3)\;. \label{frog}
\end{eqnarray}
To expose the further role of gauge invariance in our result, we give the explicit form of the pair production part ${\tilde\Delta}^\mu$ (from which one may write down ${\tilde\Gamma}^\mu$ by crossing symmetry):
\be\begin{split}
	 \tilde\Delta^\mu(s) = &\bar{u}_{p_3}\bigg[\gamma^\mu\, B_0(s) - \frac{a^2m^2k^\mu}{2\sp{k}{p_3}\,\sp{k}{p_4}}\,\slashed k\, B_3(s) \\
	 &+ \sum_{j=1}^2 \frac{e}{2}\bigg(\frac{\slashed{a^j}\slashed{k}\gamma^\mu}{\sp{k}{p_3}} - \frac{\gamma^\mu\slashed{k}\slashed{a^j}}{\sp{k}{p_4}}\bigg)B_j(s)\bigg]v_{p_4}\;.
\end{split}
\ee
All dependence on the parameter $s$ (a `photon number') is contained in four functions $B_0\ldots B_3$. For $j=1,2,3$, and writing $f_3\equiv f_1^2 + f_2^2$, these are defined by (sum over $n=1,2,3$)
\be\label{b-defs}
	B_j(s) := \int\!\ud z\ f_j(z) \exp\bigg[\ i s z + i\alpha_n \int\limits_{z}^\infty\!\ud w\  f_n(w)\,\bigg]\;,
\ee
where the coefficients $\alpha_n$ may be read off from (\ref{PhaseSpin})--(\ref{frog}):
\be\begin{split}
	\alpha_n &= e a^n\cdot\bigg( \frac{p_4}{\sp{k}{p_4}}-\frac{p_3}{\sp{k}{p_3}}\bigg)\quad n=1,2\;, \\
	\alpha_3 &= -\frac{a^2m^2}{2}\bigg(\frac{1}{\sp{k}{p_4}}+\frac{1}{\sp{k}{p_3}}\bigg)\;.
\end{split}
\ee
The functions (\ref{b-defs}) are finite, provided the $f_j$ vanish asymptotically, as is the case for pulsed fields. We assume this behaviour from here on. (The periodic plane wave case may be recovered in a suitable limit, see \cite{Kibble:1965zza}.)  The fourth function $B_0$ is defined as in (\ref{b-defs}) but without any damping factor of $f$ under the integral. Consequently, $B_0$ is the Fourier transform of a pure phase, and some prescription is required for calculating it, at least numerically. Regularisations of this integral (or its equivalent in the nonlinear Compton part of the amplitude) were suggested in \cite{Boca:2009zz, Seipt:2010ya}, but there is a more fundamental method of defining $B_0$: one appeals to gauge invariance. Making a (quantum) gauge transformation in (\ref{S1}), one finds that $S_{fi}$ is gauge invariant provided that (sum over $n=1,2,3,$)
\be\label{FromGauge}
	s\, B_0(s) = \alpha_n B_n(s)\;, 
\ee
and similarly for analogous functions in $\Gamma^\mu$. Hence, by defining $B_0$ in terms of the well-behaved $B_j$, gauge invariance reduces the number of functions in play from four to three. This result should be compared with equations (A3) in \cite{Nikishov:1964zza}, (6) in \cite{Nikishov:1964zz},  and (23) in \cite{Mackenroth:2010jr}: those expressions were obtained either as identities of special functions or by regularisation, but they are unified and explained physically by gauge invariance.

One now has all the necessary  ingredients to calculate (\ref{full}). To obtain the full rate, one takes $|S_{fi}|^2$ and divides out the volume transverse to the pulse, as usual. The remaining steps are to sum/average over spins and perform the final state integrations, where again it is natural to use lightfront variables. There are nine integrals, corresponding to three momentum components for three outgoing particles. Three of these are eliminated by the remaining $\delta^\mathsf{lf}$. Writing $K$ for the bracketed term in (\ref{full}), one obtains, with $\alpha\equiv e^2/(4\pi)$, 
\be\label{R}
	R = \frac{\alpha^2}{4\omega^2}\left[\prod\limits_{p_3,p_4}\int\!\frac{\ud^2p_\LCperp}{(2\pi)^3}\int\limits_0^\infty\!\frac{\ud p_\LCm}{2p_\LCm}\right] \frac{ \theta({p_2}_\LCm)}{{p_2}_\LCm}\sum\limits_{\text{spins}}|K|^2\bigg|_{\text{shell}} \;.
\ee
The instruction `shell' indicates that each $p_\LCp$ is evaluated on shell, i.e. $p_\LCp =(p_\LCperp^2+m^2)/(4p_\LCm)$, and $p_2$, which is eliminated by momentum conservation, obeys ${p_2}_\LCperp=(p_1-p_3-p_4)_\LCperp$ and ${p_2}_\LCm=(p_1-p_3-p_4)_\LCm$. The remaining integrals are over the momenta of the produced electron-positron pair. The full emission probability is a function of the laser amplitude, $a$, the pulse geometry, and the incoming momenta which appear through $\sp{k}{p_1}$. This completes the calculation.

Despite the length of the expressions involved, our final result for the rate is not more complicated than that of the periodic plane wave case studied in \cite{Hu:2010ye}. Let us compare the expressions therein with our own results. There are three real differences. The first is that the discrete sums over photon number are replaced by Fourier integrals over $r$ and $s$. The second is that the shifted mass plays, in general, no role in our expressions. Finally, our S-Matrix contains two distinct terms corresponding to the one- and two-step processes. 

An important step in the calculation of the rate is the evaluation of the functions (\ref{b-defs}). In general, these must be calculated numerically. In various limits, though, analytical approximations exist. The weak field limit is an obvious example: one simply expands everything in powers of the field amplitude $a$. What is more interesting for modern experiments is the high intensity limit, which amounts to $a\gg 1$. This is studied in \cite{Mackenroth:2010jr} in the context of nonlinear Compton scattering, using an asymptotic expansion of the (equivalents of) the $B_j$. Let us adapt this to our trident process: the idea is to look for points of stationary phase in (\ref{b-defs}). However, one can show from the kinematics that $s- \alpha_j f_j(\sp{k}{x})\not =0\, \forall\, \sp{k}{x}$ and hence no points of stationary phase exist. Thus, one can try to instead deform the contours into the complex plane and use the method of steepest descent. This is a well understood technique and may be applied immediately to the trident amplitude. It would be interesting to try and understand the physics behind this approximation in more detail: as a first step, one could return to the infinite plane wave case and attempt to establish a connection between the saddle points in the complex plane and the photon number, or the effective electron mass. 

In conclusion, we have given the first full calculation of trident pair production in a laser field, using strong-field QED. We have included finite size effects due to the ultra-short duration of modern pulses. Both the one-step and two-step processes involved have been explicitly identified, and our results are in agreement with the optical theorem. We have also revealed the role gauge invariance plays, not only in the trident process, but also in nonlinear Compton scattering and stimulated pair production, through the relation (\ref{FromGauge}). We remark that our approach is equally valid for laser assisted M\o ller scattering, as this is just the crossed process of trident pair production: the appropriate S-Matrix element is obtained from (\ref{full}) by taking the outgoing positron to be an incoming electron. 

Contrary to recent claims in the literature, we have shown that there is no divergence in the trident amplitude.  Despite this, the numerical methods previously employed to calculate the amplitude, from which a great deal of information was obtained, are equally appropriate here. In particular, one may now easily compare the contributions of the one-step and two-step processes, as well as test the old Weizs\"acker-Williams approximation for the former \cite{Bula:1997eh}. It will be extremely interesting to see what these investigations reveal.  

A.~I. thanks Thomas Heinzl, Martin Lavelle and Mattias Marklund for useful discussions.
A.~I. is supported by the European Research Council under Contract No. 204059-QPQV.


\begin{thebibliography}{99}
\bibitem{gauge}
{The gauge fixing parameter is $\xi=1$ in Feynman gauge.}
\end{thebibliography}
\end{document}